\documentclass[pra,showpacs,twocolumn,longbibliography,secnumarabic]{revtex4-2}
\usepackage{graphicx}
\usepackage{amssymb}
\usepackage{float}
\usepackage[T1]{fontenc}
\usepackage{amsmath}
\begin{document}
\title
{On-demand continuous-variable quantum entanglement source for integrated circuits}
\author{Mehmet G\"{u}nay$^{\dagger,1}$}
\author{Priyam Das$^{\dagger,2}$}
\author{Emre Yuce$^{3}$}
\author{Emre Ozan Polat$^{4}$}
\author{Alpan Bek$^{3}$}
\author{Mehmet Emre Tasgin$^{5}$}
\affiliation{$^\dagger$Contributed equally}
\affiliation{ $^{1}$Department of Nanoscience and Nanotechnology, Faculty of Arts and Science, Mehmet Akif Ersoy University, 15030 Burdur, Turkey}
\affiliation{ $^{2}$Department of Physics, Bankura Sammilani College, Kenduadihi, Bankura, WB-722101, India}
\affiliation{ $^{3}$Department of Physics, Middle East Technical University, 06100 Ankara, Turkey}
\affiliation{ $^{4}$Faculty of Engineering and Natural Sciences, Kadir Has University, Cibali, Istanbul 34083, Turkey}
\affiliation{ $^{5}$Institute  of  Nuclear  Sciences, Hacettepe University, 06800 Ankara, Turkey }
\begin{abstract}
Integration of devices generating nonclassical states~(such as entanglement) into photonic circuits is one of the major goals in achieving integrated quantum circuits~(IQCs). This is demonstrated successfully in recent decades. Controlling the nonclassicality generation in these micron-scale devices is also crucial for robust operation of the IQCs. Here, we propose a micron-scale quantum entanglement device whose nonlinearity (so the generated nonclassicality) can be tuned by several orders of magnitude via an \textit{applied voltage} without altering the linear response. Quantum emitters~(QEs), whose level-spacing can be tuned by voltage, are embedded into the hotspot of a metal nanostructure~(MNS). QE-MNS coupling introduces a Fano resonance in the ``nonlinear response''. Nonlinearity, already enhanced extremely due to localization, can be controlled by the QEs' level-spacing. Nonlinearity can either be suppressed (also when probe is on the device) or be further enhanced by several orders. Fano resonance takes place in a relatively narrow frequency window so that $\sim$meV voltage-tunability for QEs becomes sufficient for a \textit{continuous} turning on/off of the nonclassicality. This provides as much as 5 orders of magnitude modulation depths.
\end{abstract}
\maketitle

\section{Introduction}

Quantum optics has been revolutionizing computational power that led to  recent demonstration of quantum advantage~\cite{zhong2020quantum,harrow2017quantum}. 
This exciting development accompanied by quantum networks~\cite{kimble2008quantum}, utilizing quantum teleportation~\cite{bennett1993teleporting,jiang2020quantum}, will surely shape the future. 
The instant data distribution to many parties can replace the classical interconnects~\cite{awschalom2021development} in conventional processors which limit the operation frequencies~\cite{ozbay2006plasmonics}.  
While these exciting demonstrations in quantum optics are encouraging, the field is facing major challenges in developing  scalable integration of components and most importantly on-demand quantum sources for diminishing errors in computation.

The profound efforts on the scalable integration of quantum optics ---integrated quantum circuits--- which aim to enable operation of quantum computation and quantum communication on a single medium retains a great excitement~\cite{elshaari2020hybrid, andersen2015hybrid}. 
Single-photon sources~\cite{wang2017high,PhysRevLett.118.130503,rodt2021integrated,ostfeldt2022demand} and/or continuous-variable sources are required to implement a scalable quantum computation scheme. 
Remarkably, entanglement swapping between discrete-variable and continuous-variable optical system indicate a connected nature of the two systems~\cite{takeda2015entanglement}.  
The latter, however, has an overwhelming advantage: it provides stability and repeatable manufacturability given the well established modern lithographic infrastructure and already existing integrated optical architectures~\cite{vaidya2020broadband}. 

Integration of a scalable quantum source necessitates the controlled generation and manipulation of entangled and/or squeezed light at much smaller (micron-scale) dimensions. 
This stimulated intense research on the micron-scale generation of quantum states, for instance, quadrature-squeezed states on silicon-nitrite chips~\cite{vaidya2020broadband} (continuous-variable entanglement~\cite{braunstein2005quantum,bowen2003experimental,masada2015continuous}). Nonlinear interactions on chip is the key for generating an entangled continuous-variable source. 
Nonlinear frequency conversion rate is either fixed~\cite{zhao2020near}; or it can be controlled by auxiliary light~\cite{lu2021efficient}, by tuning the resonances~\cite{hallett2018electrical} and optical filters~\cite{foster2019tunable}~(i.e., voltage-controlled preparation of two-photon states~\cite{heindel2017bright}), and by adjusting phase-matching condition~\cite{wang2020strong}.  
The control on the production (switching) of quantum nonclassicality in a circuit will provide the key control on a quantum source which can be integrated on a chip.

In this paper, we study an integrable (micron-scale) entangler where nonclassicality generation can not only be switched on/off, but also {\it tuned continuously} by an applied voltage. 
Furthermore, one can achieve modulation depths as large as 5 orders of magnitude. 
The entanglement (nonclassicality) switch is based on the Fano-control of nonlinear response of a metal nanostructure~(MNS). Linear response is not altered.
A quantum emitter~(QE) is positioned at the hot spot of the MNS and creates the Fano resonance in the nonlinear response. 
The level-spacing of the QE can be tuned via an applied voltage~\cite{hallett2018electrical,shibata2013large,chakraborty2015voltage,schwarz2016electrically,yu2022electrically}, which also controls the rate of the nonlinear conversion ---thus, the nonclassicality of the system. 
Here, as an example, we work the squeezing and entanglement generation at the fundamental frequency~($\omega$) of a Fano-controlled second harmonic generation~(SHG) process. 
However, such a control can be achieved also in other nonlinear processes~\cite{PhysRevB.93.035410,PhysRevB.104.235407}.

Fano resonances appear at relatively sharp frequency bands~\cite{postaci2018silent,PhysRevB.101.165412}. This feature may be disadvantageous in achieving broadband nonlinearity enhancements in MNSs. Here, however, we turn the sharpness of the Fano resonances into an advantage, because a smaller voltage tuning for the QE level-spacing~($\omega_{_{\rm QE}}$), $\sim$meV~\cite{hallett2018electrical,shibata2013large,chakraborty2015voltage,schwarz2016electrically}, comes to be sufficient for turning on and off the nonclassicality.

\begin{figure}
\begin{center}
\vskip-0.3truecm
\includegraphics[width=80mm]{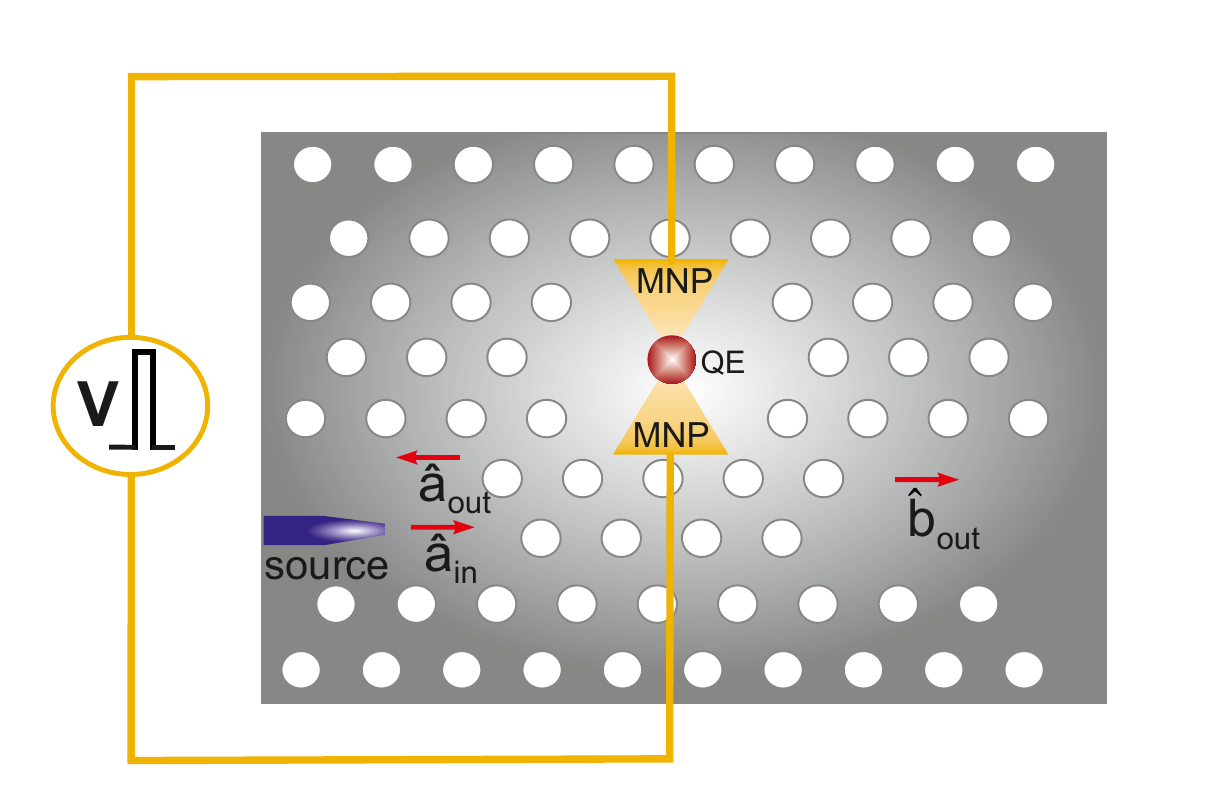} 
\vskip-0.3truecm
\caption{Micron-scale, voltage-tunable integrated entanglement device. Nonlinearity of the MNS is already extremely enhanced due to localization at the hotspot~\cite{celebrano2015mode}. QE(s) positioned to the hotspot induces a Fano resonance which can suppress (turn off) the localization-enhanced nonlinearity by several orders at $\omega_{_{\rm QE}}=2\omega$ or enhance it 10-100 times at around $\omega_{_{\rm QE}}\approx2.002$. Level-spacing~($\omega_{_{\rm QE}}$) is tuned by an applied voltage~\cite{hallett2018electrical,shibata2013large,chakraborty2015voltage,schwarz2016electrically,yu2022electrically}. $\hat{a}_{\rm in}$ is the input field~(integrated laser), $\hat{a}_{\rm out}$ and  $\hat{b}_{\rm out}$ are the output fields whose entanglement~(Fig.~\ref{fig3}a) and nonclassicality~(Fig.~\ref{fig3}b) are investigated.} 
\label{fig1}
\end{center}
\end{figure}

\begin{figure}
\begin{center}
\includegraphics[width=0.48\textwidth]{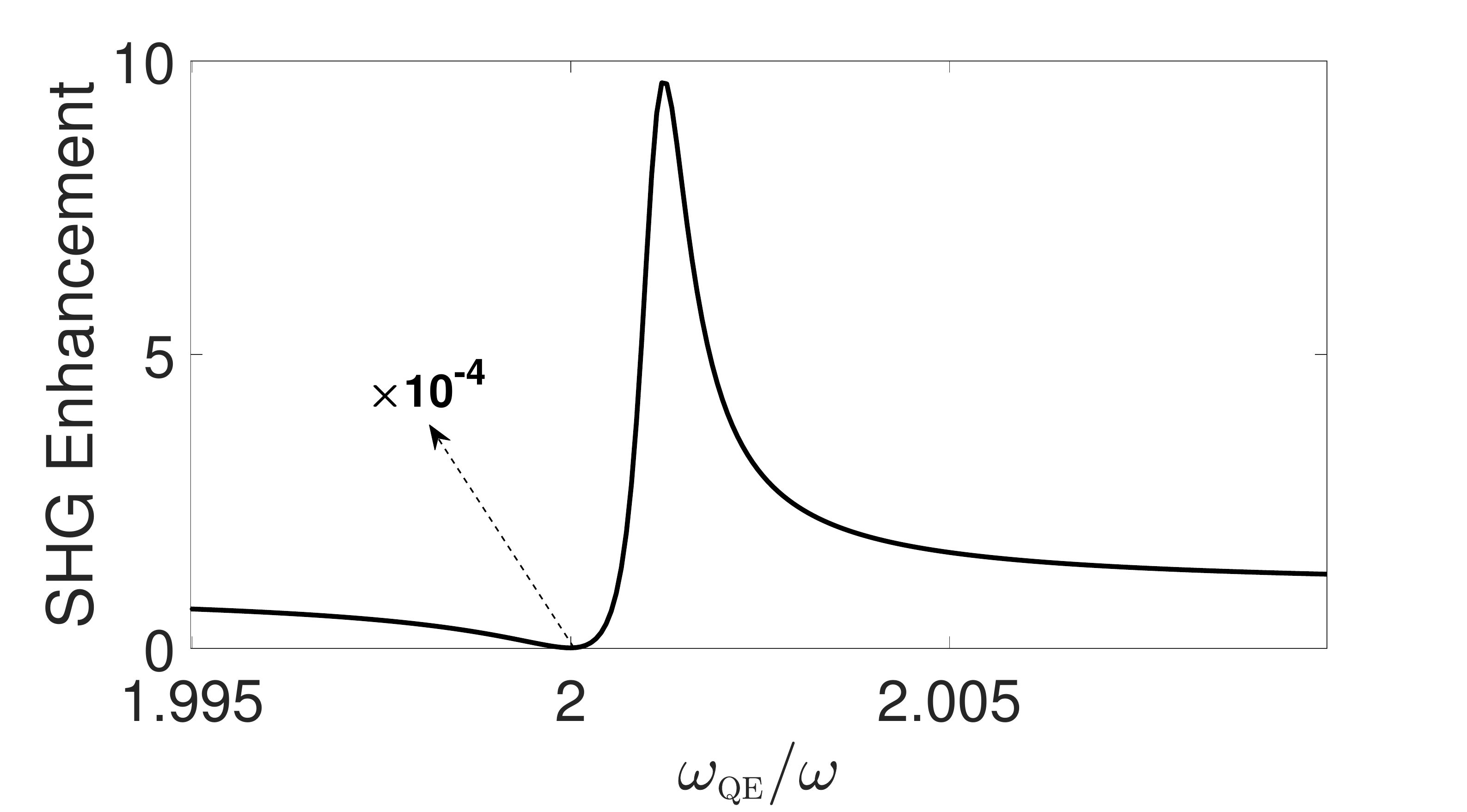}
\caption{Fano-enhancement of the SHG, which multiplies the localization enhancement~\cite{celebrano2015mode}, for different $\omega_{_{\rm QE}}$. SHG enhancement is tuned between $\times 10^{-4}$ and $\times$10 by adjusting the QE level-spacing $\omega_{\rm \scriptscriptstyle QE}$ via an applied voltage.}
\label{fig2}
\end{center}
\end{figure}

We consider a micron-sized photonic crystal cavity into which a MNS, e.g., a bow-tie antenna, is embedded, see Fig.~\ref{fig1}. The QE(s)~\footnote{One can also use more than one QE, e.g., a nanodiamond (or a 2D material) with several defect centers.}, whose level-spacing is voltage-tuned~\cite{schwarz2016electrically}, is positioned at the a few-nm-sized hotspot in the gap. A Fano resonance appears due to the MNS-QE coupling, see Fig.~\ref{fig2}. When $\omega_{_{\rm QE}}$ is tuned to $2\omega$, the SHG process (so the nonclassicality generation) is suppressed by 4 orders of magnitude, i.e., multiplied by $10^{-4}$~\cite{turkpence2014engineering,PhysRevB.93.035410,PhysRevB.104.235407}. In contrast, when the $\omega_{_{\rm QE}}$ is tuned to $\approx 2.002\omega$, the SHG is enhanced 10 times. We remark that the localization (hotspot) already enhances the SHG, for instance, by $10^6$ times~\cite{celebrano2015mode}. The pronounced Fano-suppression (enhancement) factors, $10^{-4}$ and 10, multiplies the $10^6$ localization enhancement~\cite{kamenetskii2018fano,celebrano2015mode}. The cavity is pumped by an integrated microlaser~\cite{rodt2021integrated,munnelly2017electrically} on the left hand side.

Here, we show that the SHG process taking place inside the cavity~(entanglement device) generates two kinds of nonclasssicality. (i) The two pulses emitted from the cavity in the opposite directions ($\hat{a}_{\rm out}$ and $\hat{b}_{\rm out}$ in Fig.~\ref{fig1}), at the fundamental frequency $\omega$, are entangled, see Fig.~\ref{fig3}a. (ii) The transmitted light $\hat{b}_{\rm out}$ is single-mode nonclassical, see Fig.~\ref{fig3}b. Thus, one can either (i) use the two entangled light beams or (ii) create entanglement (on the right hand side) from the nonclassicality inherited in $\hat{b}_{\rm out}$ using an integrated beam-splitter~\cite{PhysRevA.74.032333}. Beyond the generation of nonclassicality~\cite{tasgin2020measuring,PhysRevA.101.062316}, the most important thing our device can provide is the continuous tunability of the quantum nonclassicality (by several orders of magnitude) via an applied voltage with an unaltered linear response.

\section{Control of Quantumness by voltage}

\subsection{Cavity System and Hamiltonian}

Dynamics of the voltage-controlled entanglement device can be described as follows. An integrated laser of frequency $\omega$ excites the fundamental cavity mode~($ \hat{c}_1 $) on the left hand side, $\hat{\cal H}_{\rm \scriptscriptstyle  L}=\hbar(\varepsilon_{\rm \scriptscriptstyle L}\hat{c}_1^\dagger e^{-i\omega t}+H.c.)$. The cavity mode couples with the first (lower-energy) plasmon mode~($\hat{a}_1$, resonance $\Omega_1$) of the bow-tie MNS, $\hat{\cal H}_1=\hbar g_1 \hat{a}_1^\dagger\hat{c}_1+H.c.$. The $\hat{a}_1$ plasmon excitation is localized within the nm-sized hotspot located at the gap between the two metal nanoparticles. Orders of magnitude enhanced electromagnetic field ($\omega$), at the hotspot, yield the SHG process~\cite{hubert2007role}. Two localized excitations ($\omega$) in the $\hat{a}_1$ plasmon mode combine to produce a single 2$\omega$ plasmon in the second (higher energy) plasmon mode $\hat{a}_2$, $\hat{\cal H}_{\rm \scriptscriptstyle SHG}=\hbar \chi^{(2)} \hat{a}_2^\dagger \hat{a}_1\hat{a}_1 + H.c.$. The hotspot of the $\hat{a}_2$ mode, of resonance $\Omega_2$, is also at the center (gap). The SHG conversion takes place over the plasmons~\cite{PhysRevLett.108.136802} because overlap integral for this process is extremely larger due to the localization~\cite{kauranen2012nonlinear,kamenetskii2018fano}. As both incident~($\omega$) and converted~(2$\omega$) fields are localized, the SHG process can be enhanced as large as $10^6$ times ---localization enhancement~\cite{celebrano2015mode}. The level-spacing of the QE~($\omega_{_{\rm QE}}$) is around the second harmonic~(SH) frequency $2\omega$ and the resonance  $\Omega_2$ of the second plasmon mode $\hat{a}_2$, so that it is off-resonant to the fundamental frequency. The localized $\hat{a}_2$ plasmon mode couples with the QE(s) ($\hat{\cal H}_2=\hbar f |e\rangle\langle g| \hat{a}_2 + H.c.$) which introduces a Fano resonance in the SH conversion, see Fig.~\ref{fig2}. The SHG process can be controlled by the level-spacing of the QE(s), $\omega_{_{\rm QE}}$ which is tuned by an applied voltage~\cite{hallett2018electrical,shibata2013large,chakraborty2015voltage,schwarz2016electrically,yu2022electrically}. When voltage tunes the QE(s) to $\omega_{_{\rm QE}}= 2\omega$, the localization enhanced~(e.g., about $10^6$ times) SHG is suppressed $10^{-4}$ times. That is, the switch turns the SHG off. When $\omega_{_{\rm QE}} \approx2.002\omega$, this time the localization enhanced SHG is further multiplied by a factor of 10. 

The SHG process~($\hat{a}_2^\dagger \hat{a}_1 \hat{a}_1 + H.c.$) generates the nonclassicality~(e.g., squeezing) in the $\hat{a}_1$ plasmon mode~\cite{lugiato1983squeezed}. The nonclassicality of the $\hat{a}_1$ mode is transferred back to the fundamental cavity mode $\hat{c}_1$ via the beam splitter like interaction. Nonclassicality introduces in the $\hat{c}_1$ mode. The cavity mode couples to the left~($\hat{a}_{\rm out}$) and right~($\hat{b}_{\rm out}$) hand sides of the cavity.  $\hat{a}_{\rm out}$ and $\hat{b}_{\rm out}$ are related to the cavity field $\hat{c}_1$ via input-output relations~\cite{genes2008robust}. Because of the common interaction with the cavity mode~(entanglement swap~\cite{PhysRevA.72.042310}) and the nonclassicality of $\hat{c}_1$, the $\hat{a}_{\rm out}$ and $\hat{b}_{\rm out}$ modes get entangled, see Fig.~\ref{fig3}a. In addition to this entanglement, the nonclassicality~(squeezing) in the cavity mode is transferred also to the $\hat{b}_{\rm out}$ on the right hand side. Thus, $\hat{b}_{\rm out}$ is also single-mode nonclassical~(squeezed), see Fig.~\ref{fig3}b, as well as it is entangled with the $\hat{a}_{\rm out}$ mode. (Not all of the total nonclassicality can be converted into entanglement in beam-splitter like interactions, but some single-mode nonclassicality remains within the transferred mode~\cite{PhysRevA.92.052328}.) One does not have to use the entanglement of the $\hat{a}_{\rm out}$ and $\hat{b}_{\rm out}$ modes, but can also convert the single-mode nonclassicality of the transmitted $\hat{b}_{\rm out}$ mode into entanglement by placing an integrated beam splitter on the right hand side~\cite{PhysRevA.65.032323}.

\subsection{Langevin Equations}

Time evolution of the operators can be determined using the Heisenberg equations of motion, e.g., $\dot{\hat{a}}_1=[\hat{a}_1,\hat{\cal H}]$, as
\begin{small}
\begin{eqnarray}
 \dot{\hat{c}}_1 &=& -(\kappa_{1}+ i \omega_{1}) \hat{c}_1 - i g^{*}_1 \hat{a}_1 + \varepsilon_L e^{-i \omega t},\label{Eq:EOMa}\\
 \dot{\hat{c}}_2 &=& -(\kappa_{2}+ i \omega_{2}) \hat{c}_2 - i g^{*}_2 \hat{a}_2 ,\label{Eq:EOMa2}\\
 \dot{\hat{a}}_{1} &=& - (\gamma_{1} + i \Omega_{1}) \hat{a}_1 - i g_1 \hat{c}_1 - i 2  \chi^{(2)} \hat{a}^{\dagger}_{1} \hat{a}_2,\label{Eq:EOMb}\\
 \dot{\hat{a}}_{2} &=& - (\gamma_{2} + i \Omega_{2}) \hat{a}_2 - i g_2 \hat{c}_2-  i \chi^{(2)} \hat{a}^{2}_{1}- i f \hat{\rho}_{ge},\label{Eq:EOMc}\\
 \dot{\hat{\rho}}_{ge}&=&  - (\gamma_{eg}  + i \omega_{_{\rm QE}})  \hat{\rho}_{ge} + i f\hat{a}_2 (\hat{\rho}_{ee}-\hat{\rho}_gg),\label{Eq:EOMd}\\
 \dot{\hat{\rho}}_{ee} &=&  - \gamma_{ee} \hat{\rho}_{ee}  + i 2 f (\hat{a}^\dagger_2\hat{\rho}_{eg} - H.c.)],\label{Eq:EOMe}
\end{eqnarray}\end{small}
where $ \kappa_{1,2}$ and $ \gamma_{1,2} $  are the decay rates for the cavity and plasmon modes. $ \gamma_{ee} $, and $ \gamma_{eg}$ are the diagonal and off-diagonal decay rates of the QE(s). Please see Supp. Mat.~\cite{suppl} for details.


\subsection{Fano Control}

One can find the field amplitudes of the coupled cavity-MNS-QE(s) system examining the expectations of the operators, e.g., $\alpha_{1,2}=\langle\hat{a}_{1,2}\rangle$. Here, $|\alpha_2|^2$ gives the number of SH converted plasmons which governs the nonclassicality of the system. The steady-state amplitude of the second harmonic plasmons~\cite{gunay2020controlling,tasgin2018fano,turkpence2014engineering}
\begin{eqnarray}
  \alpha_2 &=& \frac{i \chi^{(2)}}{\frac{|f|^2 y}{i (\omega_{_{\rm QE}} - 2 \omega) + \gamma_{eg}}  - [i (\Omega_2 - 2 \omega) + \gamma_2]} \alpha^2_1  \label{eq:steady} 
\end{eqnarray}
is governed by the interference taking place in the denominator ${\cal D}(\omega)=\frac{|f|^2 y}{i (\omega_{_{\rm QE}} - 2 \omega) + \gamma_{eg}}  - [i (\Omega_2 - 2 \omega) + \gamma_2]$. When $\omega_{_{\rm QE}}=2\omega$, the first term of ${\cal D}$ becomes $|f|^2 y/\gamma_{eg}$ which turns out to be very large due to the QE's small decay rate~\cite{hallett2018electrical}, e.g., $\gamma_{eg}=10^{-6}\omega$. The typical values for MNS-QE coupling and population inversion are $f=0.1\omega$ and $y=\rho_{ee}-\rho_{gg}\sim -1$. Thus, the first term of ${\cal D}$ becomes of order $\sim 10^4\omega$ while the second term of ${\cal D}$ is less than unity~(1$\times \omega$). This greatly suppresses the SHG which is depicted in Fig.~\ref{fig2}. We remark that without the presence of the QE, the SHG of the MNS is governed by the second term of ${\cal D}$. The suppression is stronger when sharper resonance QE(s) are used.

In contrast, one can also enhance the SHG by preforming a cancellation in the denominator ${\cal D}$. By tuning $\omega_{_{\rm QE}}$ accordingly, the off-resonant expression $(\Omega_2 - 2 \omega)$ can be canceled by the imaginary part of the first term in ${\cal D}$, see $\omega_{_{\rm QE}}\approx2.002\omega$ in Fig.~\ref{fig2}. Therefore, tuning the level-spacing of the QE $\omega_{_{\rm QE}}$ about $\sim$meV one can continuously tune the SHG by 5-orders of magnitude and in particular turn on and off the nonclassicality. We utilize this phenomenon as a voltage-controlled integrable quantum entanglement device.

\begin{figure}
\begin{center}
\includegraphics[width=80mm]{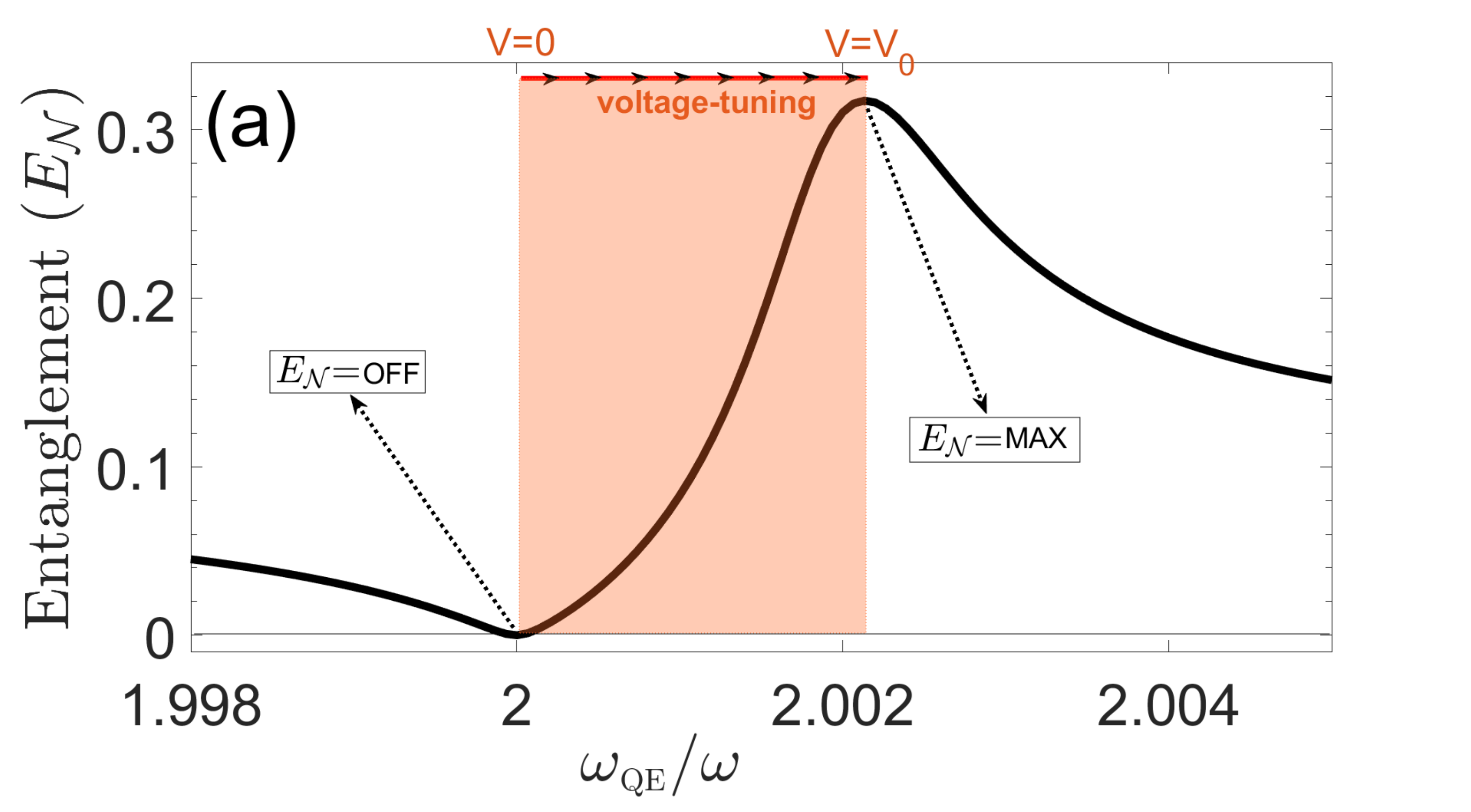}
\includegraphics[width=80mm]{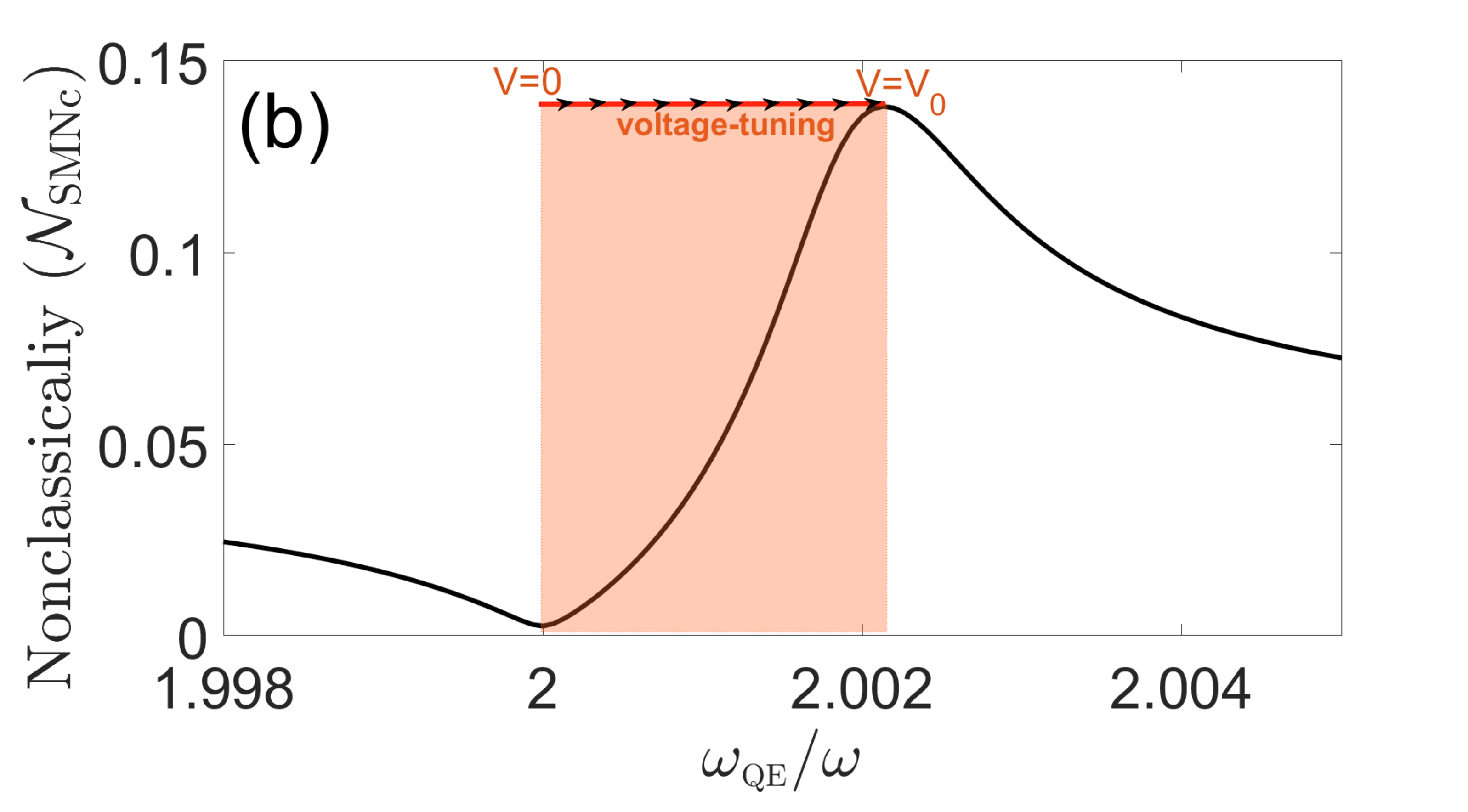}
\caption{(a) Degree of the entanglement~(log-neg) of the two output fields $\delta a_{\rm out}$ and $\delta b_{\rm out}$ in Fig.~\ref{fig1}. Voltage-tuning of the QE(s) level-spacing between $\omega_{_{\rm QE}}=2\omega$ and $\omega_{_{\rm QE}}\approx2.002\omega$ turns off and on the entanglement, respectively. Thus an undesired pulse is not turned into nonclassical. (b) Single-mode nonclassicality~(${\cal N}_{\rm SMNc}$) of the output mode $\hat{b}_{\rm out}$ in units of log-neg~\cite{asboth2005computable,tasgin2020measuring}. The right-going output can be used to generate entanglement via an integrated beam-splitter~\cite{PhysRevA.65.032323}.}
\label{fig3}
\end{center}
\end{figure}

\subsection{Generation of Nonclassicality and Entanglement} 

We calculate the nonclassicality of the system using the standard (quantum noise) methods~\cite{genes2008robust}. The quantum nonclassicality features of a system is determined solely by the fluctuations~\cite{PhysRevA.49.1567}~(noise, e.g., $\delta \hat{a}_1$) around the expectations values of the fields~($\alpha_1=\langle \hat{a}_1\rangle$), i.e., $\hat{a}_1=\alpha_1+\delta \hat{a}_1$ and $\hat{c}_1=\beta_1+\delta \hat{c}_1$. The Langevin equations for the noise operators can be written as
\begin{subequations}
\begin{eqnarray}
  \delta \dot{\hat{c}}_1 &=&  - [\kappa_1 + i(\omega_1-\omega) \delta \hat{c}_1 
   - i g_1^{*} \delta \hat{a}_1+\delta \hat{c}_{in}^{(1)}, \label{fleqa_lin}
   \\
 \delta \dot{\hat{c}}_2 &=&  - [\kappa_2 + i(\omega_2-2\omega) \delta \hat{c}_2 
   - i g_2^{*} \delta \hat{a}_2+\delta \hat{c}_{in}^{(2)}, \label{fleqa_lin}
   \\
  \delta \dot{\hat{a}}_1 &=&  - [\Gamma_{1} + i (\Omega_1-\omega)] \delta \hat{{a}}_1 - i g_1 \delta \hat{c}_1 \nonumber \\
   &-&  2 i \chi^{(2)} (\alpha^{*}_{1} \delta \hat{a}_2 + \alpha_{2} \delta \hat{a}_1 ) \label{fleqb_lin},\\
 \delta \dot{\hat{a}}_2 &=&  - [\Gamma_{2} + i (\Omega_2-2\omega)] \delta \hat{a}_2 - i g_2 \delta \hat{c}_2 \nonumber \\
  &-&i \chi^{(2)} (2 \alpha_1 \delta \hat{a}_1 ).
  \label{fleqc_lin} 
\end{eqnarray}
\end{subequations}
Quantum optics experiments with MNSs~\cite{fasel2006quantum,huck2009demonstration,varro2011hanbury,di2012quantum,krenn1999squeezing} demonstrate that, intriguingly, plasmon excitations can hold entanglement much longer times~\cite{PhysRevLett.94.110501,tame2013quantum,fasel2006quantum,huck2009demonstration,varro2011hanbury,di2012quantum} compared to their decay rates controlling field amplitudes. For this reason, empirically, we need to consider a smaller decoherence rate $\Gamma_{1,2}$ for the noise operators $\delta\hat{a}_{1,2}$. (We also present our results with $\Gamma_{1,2}=\gamma_{1,2}$ in the Supplementary Material~(SM)~\cite{suppl} for convenience.)

First, we calculate the entanglement between the reflected~($\hat{a}_{\rm out}$) and transmitted~($\hat{b}_{\rm out}$) pulses, see Fig.~\ref{fig1}. We calculate the logarithmic-negativity~($E_{\cal N}$) which is an entanglement measure for Gaussian states~\cite{PhysRevLett.95.090503}. As we use the standard linearization method~\cite{genes2008robust} for calculating the evolution of the noise operators, the fields stay Gaussian and $E_{\cal N}$ can be employed as a measure. In Fig.~\ref{fig3}a, we observe that the log-neg can be tuned continuously on and off between $\omega_{_{\rm QE}}=2\omega$ and $\omega_{_{\rm QE}}\approx2.002\omega$.

Next, we also calculate the single-mode nonclassicality~(${\cal N}_{\rm \scriptscriptstyle SMNc}$) of the transmitted wave $\hat{b}_{\rm out}$. We employ the entanglement potential~\cite{PhysRevLett.94.173602} as measure for the single-mode nonclassicality. Only a nonclassical single-mode state can create entanglement at the output of a beam-splitter~\cite{PhysRevA.65.032323}. Entanglement potential is the degree of entanglement a nonclassical state creates at the beam-slitter output which can also be quantified in terms of log-neg. We remind that not all of the nonclassicality of a mode could be converted into entanglement at the beam splitter output. In Fig.~\ref{fig3}b, we observe that ${\cal N}_{\rm \scriptscriptstyle SMNc}$ can similarly be continuously tuned by several orders of magnitude via a $\sim$meV adjustment of the QE level-spacing~\cite{hallett2018electrical,shibata2013large,chakraborty2015voltage,schwarz2016electrically,yu2022electrically}. We note that energy level-spacing modulations as large as $\sim$25 meV~\cite{shibata2013large}, or even larger ones~\cite{muller2005wave,empedocles1997quantum} are observed.

Therefore, one can use the (i) entanglement of two pulses propagating in opposite directions or (ii) convert the nonclassicality of the transmitted $\hat{b}_{\rm out}$ pulse into entanglement on the right hand side using an integrated beam splitter. Such a modulation is important for generating the nonclassicality when the desired pulse is passing through the device, but turning it off for an unwanted pulse. We note that the sample setup, Fig.~\ref{fig1}, can be placed even into a smaller cavity~\cite{akahane2003high}.

\section{Experimental Implementation of the Active Entanglement Device}

The abovementioned narrow-band and tunable Fano resonance-based device architecture promises for the electrically controllable switches that paves the way towards the active control of the nonclassicality generation. Although the electrically tunable Fano resonance device has so far not been experimentally realized as a nonclassicality switch, the suggested system of  QE positioned at the hot spot of the MNS has experimentally been demonstrated to be applicable using various methodologies~\cite{lyamkina2016monolithically,santhosh2016vacuum,jiang2021single}.  Lyamkina {\it et al}. demonstrated a light-matter coupling of single InAs quantum dots monolithically integrated into electromagnetic hot-spots of sub-wavelength sized metal nanoantennas~\cite{lyamkina2016monolithically}. The authors formed self-assembled InAs quantum dots~(QDs) on a molecular beam epitaxy grown (001)~GaAs substrates and lithographically formed the bow-tie antenna structures followed by an e-beam metallization and lift-off. The antennas are reported to enhance the emission intensity of single QD by 16 times and the fabricated structure allows intrinsic electrical connectivity allowing for the Stark tuning of the QEs. 

Alternately, Santosh {\it et al}. demonstrated vacuum Rabi splitting by using the strong coupling of silver bow-tie plasmonic cavities loaded with semiconductor quantum dots~\cite{santhosh2016vacuum}. Authors used interfacial capillary forces to drive commercially available CdSe/ZnS QDs into the lithographically patterned holes in the bowtie gaps. A strong coupling rate of 120 meV has been reported with a single QD system and, by using the fabricated structures, authors demonstrated the transparency dip in the spectral measurements due to the Rabi splitting. Jiang {\it et al}. reported an automatically located QE structure by plasmonic nanoantennas bypassing the accurate nanoscale alignment of the source at the plasmonic hotspot~\cite{jiang2021single}. The authors have demonstrated that 11 nm  diameter, single CdSe/ZnS QDs could be trapped achieving a trap stiffness of 0.6 (fN/nm)/mW yielding 7 times increased brightness, 2 times shortened lifetime. 

The waveguide integration and CMOS compatibility is another experimental aspect that should be satisfied for the realization of entanglement devices yielding the production of IQCs. To that end, number of fabrication approaches have been reported that are compatible to the suggested model of entanglement~(SHG) device. Hallett {\it et al}.~\cite{hallett2018electrical} demonstrated the electrical control of resonant photon scattering from QDs that are embedded in a waveguide coupled photonic device to provide a switchable nonlinear response at the single photon level. Our proposed approach is compatible to be implemented as an embedding of QEs inside the bulk p-i-n or Schottky diode structure that could provide a fast frequency tuning of the QEs with the DC Stark effect. 

More interestingly, integrating synthetic systems such as $\pi$-conjugated molecules and colloids between the metallic antenna structures can provide the quantum-confined Stark effect that can effectively tune the QE’s wavelength due to the intrinsic charge carrier confinement in three dimensions. Although the synthetic QDs suffer from the heterogeneity that can create undesirable temporal fluctuations on a single particle level~\cite{empedocles1997quantum}, Muller {\it et al}. have demonstrated a controlled manipulation of the single particle wave function in semiconducting colloidal QDs by asymmetric growth of shell materials that yields the localization of charge carriers at specific distances from the core~\cite{muller2005wave}.  To that end, synthetic QEs could provide more scalable and low-cost SHG device fabrication routes compared to epitaxial grown structures.

Furthermore, electrical tunability of the nonclassicality switch can be limited by the screening of the gate electric fields. To overcome that, Shibata {\it et al}. have demonstrated the use of a liquid gate electrical double layer~(EDL) in gating of zero dimensional QDs allowing to tune the electronic states over a wide range that is not possible in solid state dielectric gate transistors~\cite{shibata2013large}. The authors have found that the efficiency of EDL gating (350 meV/V) is 6 times higher than the back gating (60 meV/V) for the QDs yielding around 25 meV charge addition energy between N=1 and 2. Although the ionic liquid gate technology has been demonstrated as a powerful tool to effectively shift the Fermi energy in solids~\cite{polat2013broadband,yuan2009high,shimotani2006gate}, the fast cyclic switching, and the waveguide integration remained limited due to the involvement of a liquid state material that operates via the propagation of the ionic content.  However, with the development of CMOS compatible thin film capacitors, effective gating technologies promise for the development of active SHG devices that are coupled within a waveguide structure.

The recent momentum in the van der Waals heterostructures based research has revealed the usage of defect states as QEs. Schwarz {\it et al}.~\cite{schwarz2016electrically} have demonstrated the electrically pumped sharp luminescence from individual defects in WSe2 that are sandwiched in a graphene/hBN/WSe2/hBN/graphene structure.  Our active SHG device using metallic nanoantenna structure(MNS) could be formed by using 2D materials such that MNS/local defect/MNS is formed. To achieve integrated photonics technology, waveguide coupled structures could be realized following the reported integration methods of 2D materials~\cite{muench2019waveguide,goossens2017broadband,gao2015high}.


\section{Summary}

In summary, we introduce a micron-scale quantum entanglement device which can be integrated into (quantum) photonic circuits. The nonclassicality can be supplied into the integrated quantum circuit on-demand by applying a voltage on the device. Unlike the integrable quadrature squeezing generators presented in the literature~\cite{vaidya2020broadband, Mondain, zhang2021squeezed,lenzini2018integrated}, the nonclassicality can be turned off also when a pulse is passing through the device. The voltage can also continuously tune the degree of the generated entanglement/nonclassicality. The extraordinary large modulation depth~($10^5$) results from the suppression feature of the Fano resonance~(in the nonlinear response). Moreover, the linear response of the device is not altered in the tuning~\cite{yu2022electrically,dhama2022all},~\footnote{In Ref.~\cite{yu2022electrically}, nonlinear response of a quantum emitter is tuned electrically by readjusting the energy levels of the emitter. This alters also the linear response of the material and such a large modulation depth per QE level-spacing tuning cannot be achieved.}. Although we study the tuning of nonclassicality on a second harmonic generation process here, the same method can be applied also on the the voltage-tuning of other nonlinear processes. The details of our calculations following the standard methods can be found in the SM~\cite{suppl}.

\bibliography{bibliography}

\end{document}